%% file: HIN-17-002_temp.tex
\pdfoutput=1

\documentclass[11pt,twoside,a4paper,cmspaper,final,collab]{cms-tdr}
\def\svnVersion{439014P}\def\cmsCernNoTag{CERN-EP-2017-239}\def\cmsCernDate{\today}\def\cmsMessage{Published in Physical Review Letters as \href{http://dx.doi.org/10.1103/PhysRevLett.119.242001}{\doi{10.1103/PhysRevLett.119.242001.}}}

\begin{document}\cmsNoteHeader{HIN-17-002}

\hyphenation{had-ron-i-za-tion}
\hyphenation{cal-or-i-me-ter}
\hyphenation{de-vices}
\RCS$Revision: 433393 $
\RCS$HeadURL: svn+ssh://svn.cern.ch/reps/tdr2/papers/HIN-17-002/trunk/HIN-17-002.tex $
\RCS$Id: HIN-17-002.tex 433393 2017-11-08 21:01:51Z alverson $

\newcommand{\rootsNN} {\ensuremath{\sqrt{\smash[b]{s_{_{\mathrm{NN}}}}}}\xspace}
\newcommand{\roots} {\ensuremath{\sqrt{\smash[b]{s}}}\xspace}
\newcommand{\pPb}{\ensuremath{\Pp\mathrm{Pb}}\xspace}
\newcommand{\pp}{\ensuremath{\Pp\Pp}\xspace}
\newcommand{\pn}{\ensuremath{\Pp\Pn}\xspace}
\newcommand{\ppbar}{\ensuremath{\Pp\PAp}\xspace}
\newcommand{\jj}{\ensuremath{ \mathrm{j} \mathrm{j}^{\prime}}\xspace}
\newcommand{\mjj}{\ensuremath{m_{\jj}}\xspace}
\newcommand{\mtop}{\ensuremath{m_\text{top}}\xspace}
\newcommand{\mcfm}{\textsc{mcfm}\xspace}
\newcommand{\EPOS}{\textsc{epos}\xspace}
\newcommand{\stt}{\ensuremath{\sigma_{\ttbar}}\xspace}
\newcommand{\mylumi}{\ensuremath{ 174\pm 9\nbinv}\xspace}
\providecommand{\PN}{\ensuremath{\mathrm{N}}\xspace}

\newlength\cmsFigWidth
\ifthenelse{\boolean{cms@external}}{\setlength\cmsFigWidth{0.95\columnwidth}}{\setlength\cmsFigWidth{0.85\textwidth}}
\ifthenelse{\boolean{cms@external}}{\providecommand{\cmsLeft}{top\xspace}}{\providecommand{\cmsLeft}{left\xspace}}
\ifthenelse{\boolean{cms@external}}{\providecommand{\cmsRight}{bottom\xspace}}{\providecommand{\cmsRight}{right\xspace}}
\ifthenelse{\boolean{cms@external}}{\providecommand{\EMC}{European Muon Collaboration}}{\providecommand{\EMC}{European Muon}}

\cmsNoteHeader{HIN-17-002}
\title{Observation of top quark production in proton-nucleus collisions}

\date{\today}

\abstract{
The first observation of top quark production in proton-nucleus collisions is reported using proton-lead data collected by the CMS experiment at the CERN LHC at a nucleon-nucleon center-of-mass energy of $\rootsNN = 8.16\TeV$. The measurement is performed using events with exactly one isolated electron or muon candidate and at least four jets. The data sample corresponds to an integrated luminosity of 174\nbinv. The significance of the $\ttbar$ signal against the background-only hypothesis is above five standard deviations. The measured cross section is $\sigma_{\ttbar} = 45 \pm 8\unit{nb}$, consistent with predictions from perturbative quantum chromodynamics.
}

\hypersetup{%
pdfauthor={CMS Collaboration},%
pdftitle={Observation of top quark production in proton-nucleus collisions},%
pdfsubject={CMS},%
pdfkeywords={CMS, physics, heavy ions, top quark}}

\maketitle

The top quark, the heaviest elementary particle in the standard model, has been the subject of numerous detailed studies based on data samples with large integrated luminosities in \ppbar and \pp collisions~\cite{PDG2016} accumulated at the Fermilab Tevatron and the CERN LHC, respectively.
Until recently, top quark studies remained inaccessible in nuclear collisions because of the small
integrated luminosities of the first heavy ion runs at the LHC and the low nucleon-nucleon (NN) center-of-mass
energies ($\rootsNN$) available at the BNL RHIC. This situation changed when the 2016 LHC proton-lead (\pPb) run at $\rootsNN = 8.16$\TeV produced a data set corresponding to an integrated luminosity of 174\nbinv (equivalent to 36\pbinv of nucleon-nucleon collision data).
Top quark cross sections at the LHC are dominated by
pair production via gluon-gluon fusion processes ($\cPg\cPg\to \ttbar+X$), and are
computable with great accuracy in perturbative quantum chromodynamics (QCD)~\cite{top++,mitov}.
In proton-nucleus collisions, the top quark is a novel and theoretically precise probe of the nuclear gluon density at
high virtualities $Q^2\approx m_\cPqt^2$ (where $m_\cPqt$ is the top quark mass) in the unexplored high Bjorken-$x$
region ($x\gtrsim 2m_\cPqt/\rootsNN \approx 0.05$)~\cite{dEnterria:2015mgr,dEnterria:2017dum}.
In this region,``antishadowing'' and ``EMC'' effects~\cite{Aubert:1983xm} are expected to modify the gluon density with
respect to that in the free-proton case~\cite{EPS09,EPPS16}.
The production of top quarks thus provides information on the nuclear parton distribution functions (nPDF) that is complementary to that obtained through studies of electroweak boson production. In comparison to the \PW\ and \PZ\ cases~\cite{Khachatryan:2015hha,Khachatryan:2015pzs},
top-pair cross sections are more sensitive to gluon (rather than quark) densities, at Bjorken-$x$ values about twice as large.
Novel studies of parton energy loss using top quarks in the quark-gluon plasma formed in nucleus-nucleus collisions have also been proposed~\cite{dEnterria:2015mgr,Dainese:2016gch}. A good understanding of top quark
production in proton-nucleus collisions is crucial as a baseline for these studies.

Once produced, the top quark decays promptly without hadronizing (lifetime $c\tau \approx 0.15\unit{fm}$) into a \PW\ boson plus a bottom quark,
and top quark pair events are commonly categorized according to the subsequent decay of the two \PW\ bosons.
When one \PW\ boson decays leptonically ($\ell\nu$, with $\ell=\Pe,\mu$) and the other hadronically ($\cPq\cPaq^{\prime}$),
the $\ell$+jets final state presents a typical signature of one isolated charged lepton and momentum imbalance from the unobserved neutrino in one \PW\ decay, two light quark jets from the other \PW\ decay, and two b jets from the two original top quark decays.
Such a final state features a large
branching fraction ($\approx$30\% for the \Pe+jets and $\mu$+jets channels combined, and $\approx$34\% adding also events from the
$\cPqt\to\PW\to\tau\to\Pe,\mu$ decay chain) and moderate background contamination, and thereby provides favorable conditions for the detection of \ttbar production in proton-nucleus collisions.

This Letter describes the first observation of top quark production in nuclear collisions.
The analysis is carried out with \pPb\ collisions collected by the CMS experiment at the LHC at
$\rootsNN = 8.16$\TeV, using \ttbar candidates with the event topology described above.
The $\ttbar$ cross section is extracted from a combined maximum likelihood fit of the invariant mass of the two
 light-quark jets from the \PW{} boson decay, in different categories of events
with zero, one, or at least two \cPqb-tagged jets.

The central feature of the CMS apparatus is a superconducting solenoid of 6\unit{m} internal diameter, providing a magnetic field of 3.8\unit{T}. Within the solenoid volume are a silicon pixel and strip tracker, a lead tungstate crystal electromagnetic calorimeter (ECAL), and a brass and scintillator hadron calorimeter, each composed of a barrel and two endcap sections. Forward calorimeters extend the pseudorapidity coverage provided by the barrel and endcap detectors. Muons are detected in gas-ionization chambers embedded in the steel flux-return yoke outside the solenoid.
 A more detailed description of the CMS detector, together with a definition of the coordinate system used and the relevant kinematic variables, can be found in Ref.~\cite{Chatrchyan:2008zzk}.
The event sample of \pPb collisions collected by the CMS detector in 2016 corresponds to an integrated luminosity of $\mylumi$.
The lead nuclei and protons had beam energies of 2.56\TeV and 6.5\TeV per nucleon, respectively, corresponding to a nucleon-nucleon centre-of-mass energy of $\rootsNN = 8.16$\TeV.
The direction of the proton beam was initially clockwise and was then reversed, producing two data samples of similar size.
The pseudorapidity $\eta$ is defined such as to have positive value in the direction of motion of the proton in both data samples.
The number of collisions per bunch crossing is on average 0.5 in the combined data set.

The $\Pp\PN \to \ttbar+X$ process ($\PN = \Pp,\Pn$) is simulated using the \PYTHIA~6 Monte Carlo (MC) generator~\cite{Sjostrand:2006za}
(v.6.424, tune Z2*~\cite{Chatrchyan:2013gfi,Khachatryan:2015pea})
with a mixture of \pp and \pn interactions corresponding to their ratio in \pPb collisions.
The number of MC signal events is normalized to the perturbative QCD prediction for the \ttbar\ production cross section
at next-to-next-to-leading order (NNLO) with soft gluon resummation at
next-to-next-to-leading logarithmic (NNLL) accuracy~\cite{top++,mitov}, and scaled by the Pb mass number $A = 208$.
The value of $m_{\cPqt}$ used in all simulated samples is 172.5\GeV.
Simulated samples of \PW+jets and Drell--Yan
production of charged-lepton pairs with invariant mass larger than 30\GeV are generated using \PYTHIA~6.
The MC is used solely for the purpose of efficiency measurements and validation of the functional forms used for the background distributions,
as the latter are determined {\it in situ} from the data.
All \PYTHIA signal and background samples are embedded into \pPb events generated with \EPOS-\textsc{lhc}~\cite{epos}, tuned to reproduce the global \pPb event properties experimentally measured, and reconstructed with the same analysis code as used for the data.
Because of the different energies of the proton and lead beam, the pseudorapidity for massless particles in the laboratory frame is shifted by 0.465 units in the direction of the proton beam with respect to the NN center-of-mass frame.
The kinematics of all MC-generated events are boosted to account for this effect.
Simulated samples include an emulation of the full detector response, based on \GEANTfour~\cite{geant},
with simulated alignment and calibration conditions tuned on data,
 and a realistic description of the beam spot,
i.e., of the luminous region produced by the collisions.

A two-tier trigger system selects events of interest for offline analysis~\cite{Khachatryan:2016bia}.
 This analysis is restricted to events that fired trigger paths requiring the presence of at least one muon (electron) candidate with transverse momentum (energy) $\pt > 12\GeV$ ($\et > 20\GeV$). Looser online identification criteria are applied as compared to the offline selection, and no requirement on additional analysis objects is imposed at this level.

Particle candidates are reconstructed offline with the CMS particle-flow (PF)
algorithm~\cite{Sirunyan:2017ulk}, which
identifies and provides a list of particles using an optimized combination of information from the various elements of the CMS detector.
Events are required to contain exactly one muon~\cite{muid} or electron~\cite{emid} candidate, with $\pt > 30\GeV$ and $\abs{\eta} < 2.1$, excluding in the electron case the transition region $1.444 < \abs{\eta} < 1.566$ between the ECAL barrel and endcap, where the reconstruction of electron objects is less efficient.
The muon and electron candidates are required to be isolated from nearby hadronic activity within a cone of $\Delta R = 0.3$
around the direction of the track at the primary event vertex.
The cone is defined as
$\Delta R = \sqrt{\smash[b]{(\Delta \eta)^2 + (\Delta \phi)^2}}$, and $\Delta
\eta$ and $\Delta \phi$ are the separations in pseudorapidity and
azimuthal angle.
The scalar \pt sum of all PF candidates consistent with arising from the primary event vertex and contained within the cone of radius $\Delta R$,
excluding the contribution from the lepton candidate, is used to define a relative isolation variable,
$I_\text{rel}$, through the ratio of this sum to the $\pt$ of the lepton candidate.
A charged lepton is selected if its relative isolation discriminant value satisfies $I_\text{rel}<0.15$ (muon),
0.07 (electron in the barrel), or 0.08 (electron in one of the endcaps). These thresholds have been optimized to reduce the contamination from nonprompt leptons. To remove the Drell--Yan background, events are rejected from the analysis if they contain extra electrons (muons) that are reconstructed
using a looser set of identification criteria and have $\pt>20\,(15)$\GeV within $\abs{\eta} < 2.5\,(2.4)$.
The efficiency of the lepton selection is measured using a ``tag-and-probe'' method~\cite{inclusWZ3pb} in events enriched with \Z boson candidates
and selected by the same trigger requirements as the signal candidate events.
The combined reconstruction, lepton identification, and trigger efficiency is determined as a function of lepton \pt and $\eta$.

Events are required to have at least four reconstructed jets with $\pt> 25\GeV$ and $\abs{\eta}< 2.5$,
that are separated by at least $\Delta R = 0.3$ from the selected muon or electron.
Jets are reconstructed from the PF candidates using the anti-\kt clustering algorithm~\cite{antikt} with a distance parameter of 0.4.
Jet energy corrections extracted from the full detector simulation are applied as functions of jet \pt and
$\eta$~\cite{Chatrchyan:2011ds,JES8TeVPUB} to both data and simulated samples. A residual correction to the data is applied to account for a
small data-MC discrepancy in the jet energy response.
Jets from \cPqb\ quarks are tagged based on the presence of a secondary vertex from \PB-hadron decays, identified using
a multivariate algorithm combining tracking information~\cite{Chatrchyan:2012jua}.
The distinct \ttbar signature of two \cPqb\ jets in the event, which rarely occurs in background processes such as \PW+jets and QCD multijet
(collectively labeled as ``non-top'' background), is used to extract the signal.
The number of jets passing a threshold on the \cPqb-jet identification discriminant, corresponding to a \cPqb\ tagging efficiency of
approximately 70\% with a misidentification rate of less than 0.1\% for light-flavor jets, as estimated in simulated \pPb events, is used to classify the selected events
into no (0\cPqb), exactly one (1\cPqb{}), or at least two (2\cPqb{}) tagged-jet categories. All three event categories are exploited in a
maximum-likelihood fit in order to extract the signal cross section, and simultaneously constrain the background contamination
and determine the efficiency of the \cPqb\ jet identification.

In the $\ell$+jets final state, two light-flavor jets (\jj) are produced in the decay of one of the \PW{} bosons,
and the resonant nature of their invariant mass provides a distinctive feature of the \ttbar\ signal with respect to the main backgrounds.
Given that these light-flavor jets are correlated at production, they are also closer in phase space relative
to other dijet combinations in the event.
In cases where more than two non-\cPqb-tagged jets are found, the \jj pair with smallest separation
in the $\eta$--$\phi$ plane is used to form a \PW{} boson candidate.
The invariant mass of those two jets, \mjj, is used as input for the maximum-likelihood fit.

The parametrization of the signal in the fit model is derived from the MC simulation, while that of the backgrounds is
obtained from control regions in the data.
In the MC simulation, pairs of jets that are geometrically matched at the parton level with the light quarks coming from
$\PW \to \cPq\cPaq^{\prime}$ are marked as ``correctly-assigned'' pairs. The \ttbar signal includes correct and wrong assignments.
For all mass variables and \cPqb\ jet multiplicity categories, the \mjj spectrum is modeled for correct and wrong assignments, respectively,
by a Crystal Ball function~\cite{SLAC-R-236} summed with a gamma function,
and by a bifurcated Gaussian (i.e., a Gaussian with different widths to the left and right of its mean) summed with a Landau function~\cite{roofit}.

The background contribution from \PW+jets is assumed to be described by a Landau function, as supported by the agreement observed between the
MC simulation and a Landau parametrization in events with no \cPqb-tagged jets.
The background from QCD multijet events due to misidentified or nonprompt leptons satisfying the selection criteria is modeled with the help of dedicated background control regions. In the muon channel, the background region is selected by an inverted isolation requirement, $I_\text{rel}>0.2$, while all other selection criteria remain unchanged.
In the electron channel, the background shape is modeled with electrons that fail a looser identification requirement.
In both cases the shape of the \mjj distribution for this background is estimated with a nonparametric kernel approach~\cite{Cranmer:2000du}.
This approach is validated using events with no \cPqb-tagged jets and with missing transverse momentum (defined as the
 negative of the vectorial \pt sum of all identified particles) smaller than 20\GeV in magnitude.
The initial normalization of the QCD multijet backgrounds in the other
\cPqb{}-jet multiplicity categories is also determined from events with missing transverse momentum smaller than 20\GeV.

\begin{figure*}[htp]
\centering
\includegraphics[width=0.32\textwidth]{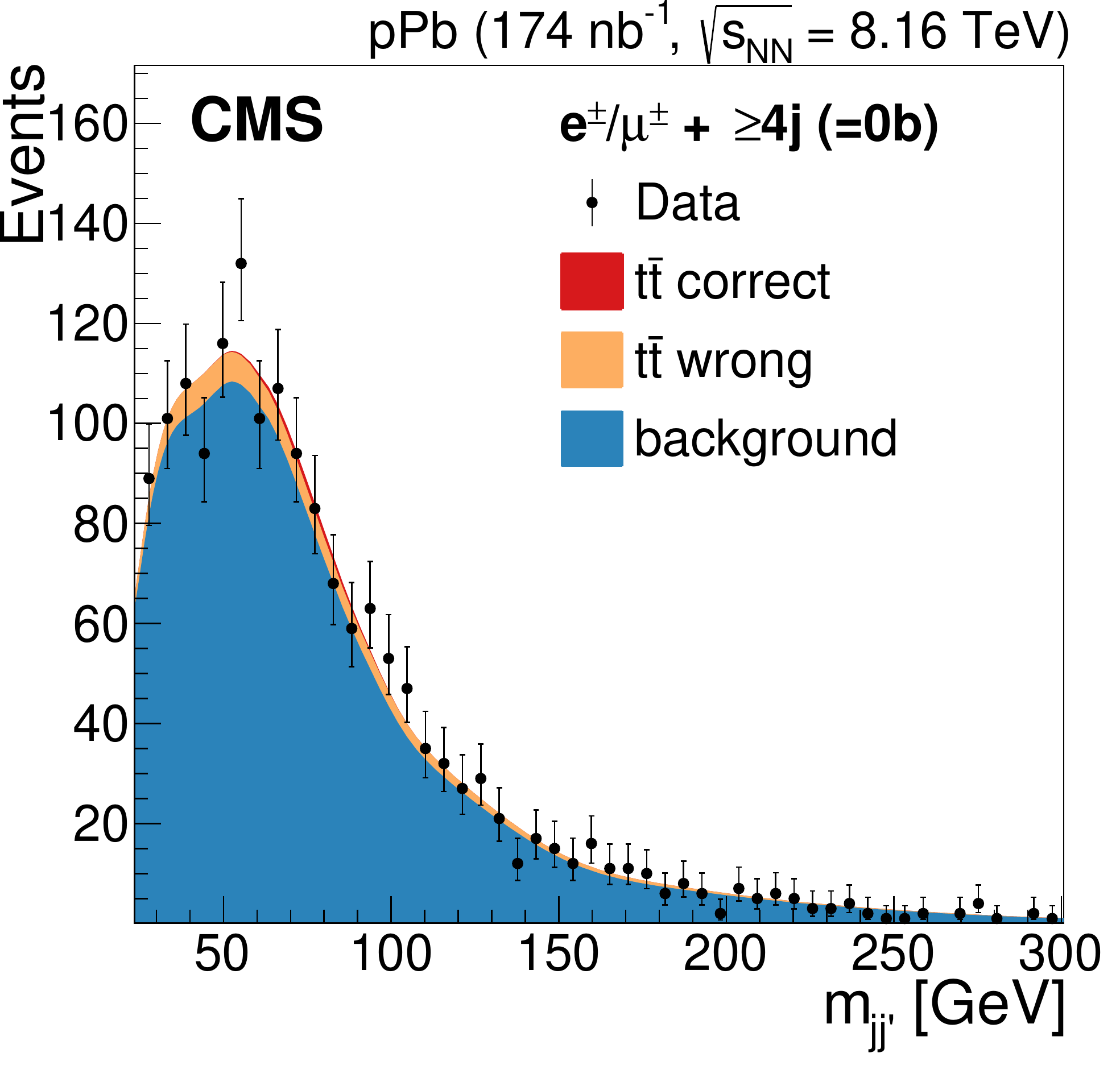}
\includegraphics[width=0.32\textwidth]{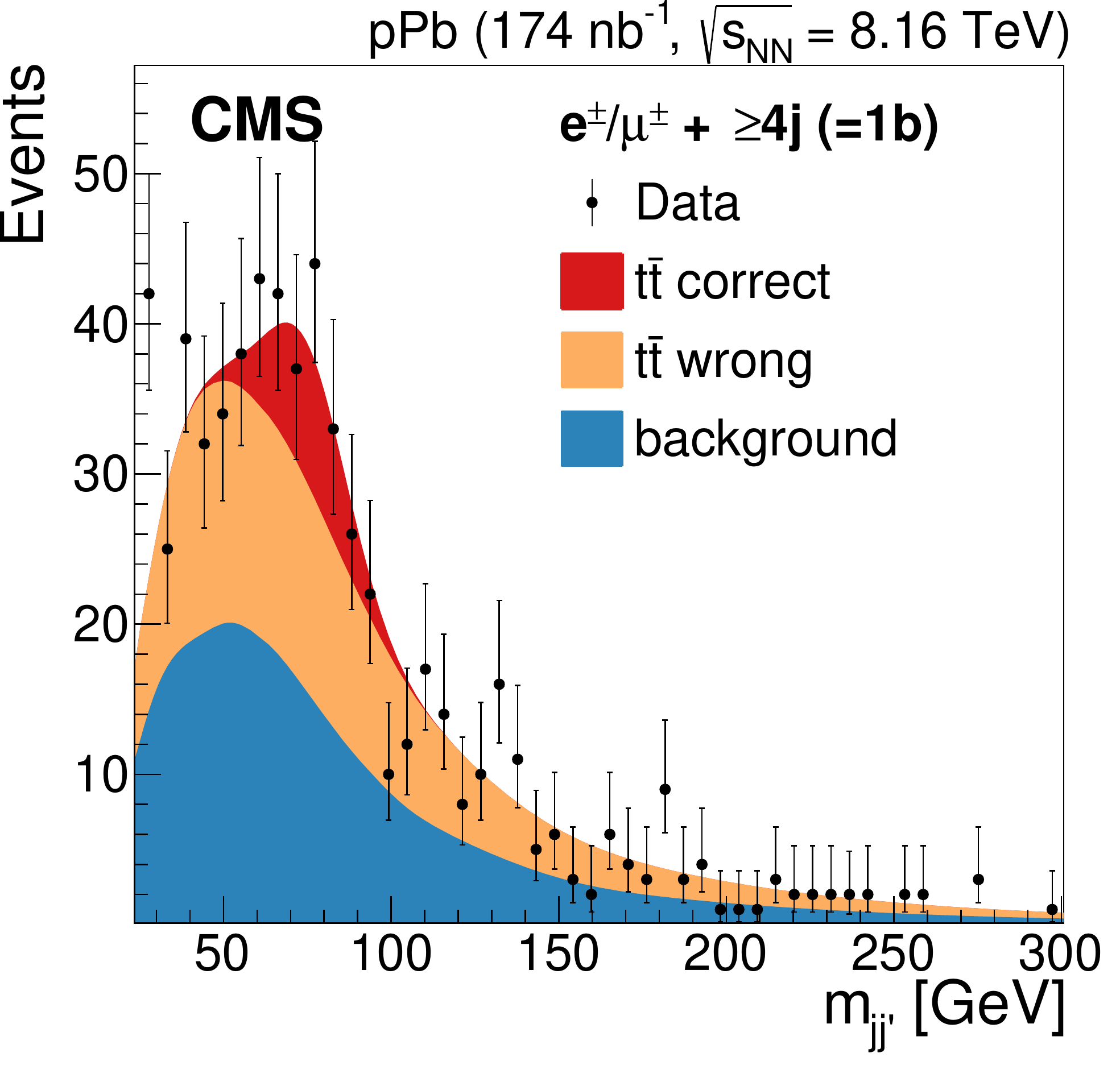}
\includegraphics[width=0.32\textwidth]{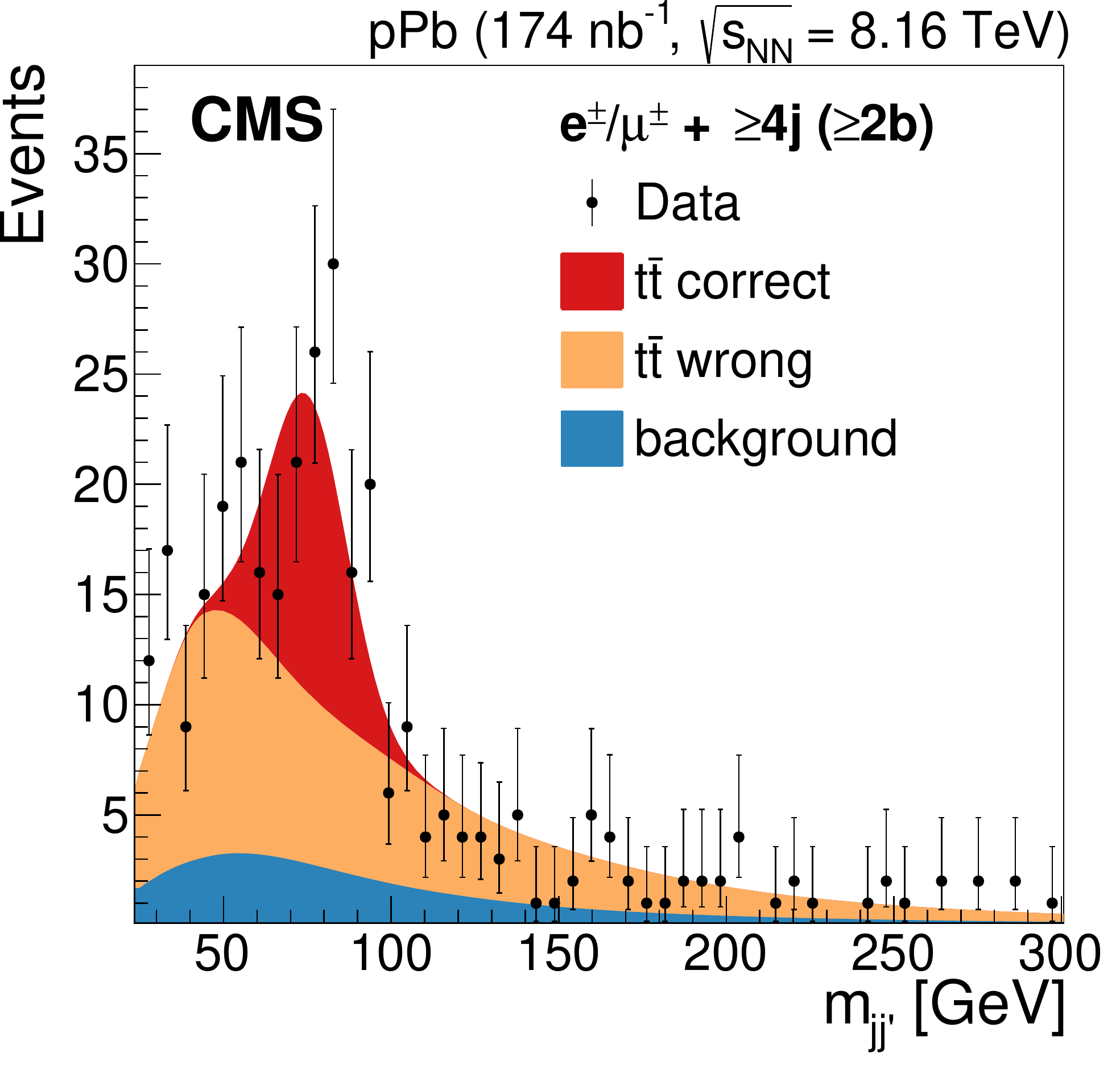}
\caption{Invariant mass distributions of the \PW\ candidate, \mjj,
in the 0 (left), 1 (center), and 2 (right) \cPqb{}-tagged jet categories after all selections.
The red and orange areas correspond to the signal simulation (correct and wrong assignments, respectively)
while the blue one corresponds to the estimated non-top background contributions.
The error bars indicate the statistical uncertainties.}
\label{fig:fitres}
\end{figure*}

The number of events in each \cPqb\ jet category is obtained by fitting the sum of the contributions for signal and backgrounds.
The free parameters of the fit are the normalization of the signal, QCD multijet, and \PW+jets yields
(as well as the parameters of their functional forms described above), the \cPqb-finding efficiency,
\ie, the probability that a jet originating from the \cPqb\ quark from a top quark decay passes both the kinematic and the \cPqb\ tagging selections,
and an overall jet energy scale factor.
Figure~\ref{fig:fitres} shows the \mjj distribution for events with zero, one, or at least two \cPqb-tagged jets, compared with the fit results.

\begin{figure*}[htp]
\centering
\includegraphics[width=0.32\textwidth]{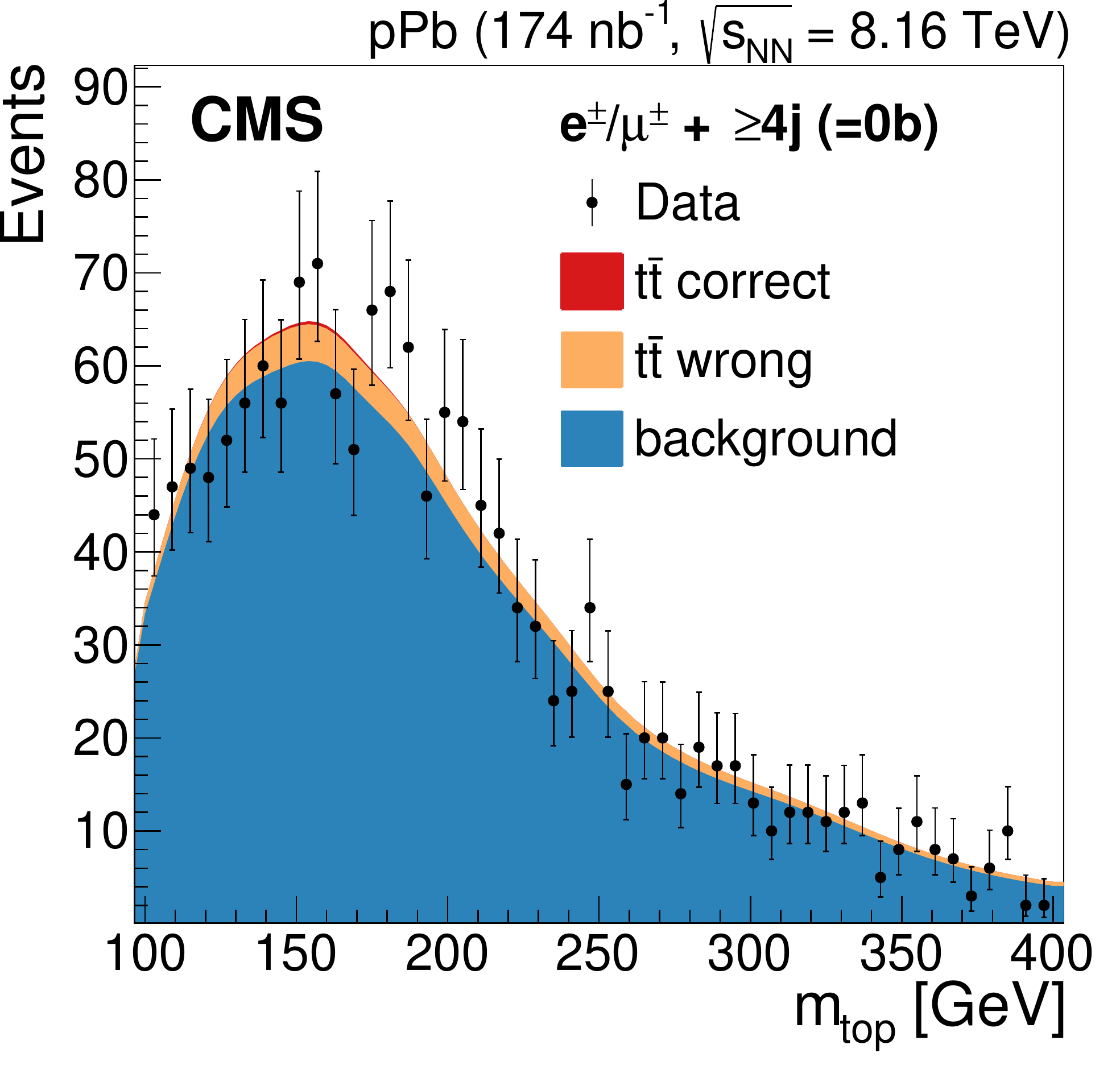}
\includegraphics[width=0.32\textwidth]{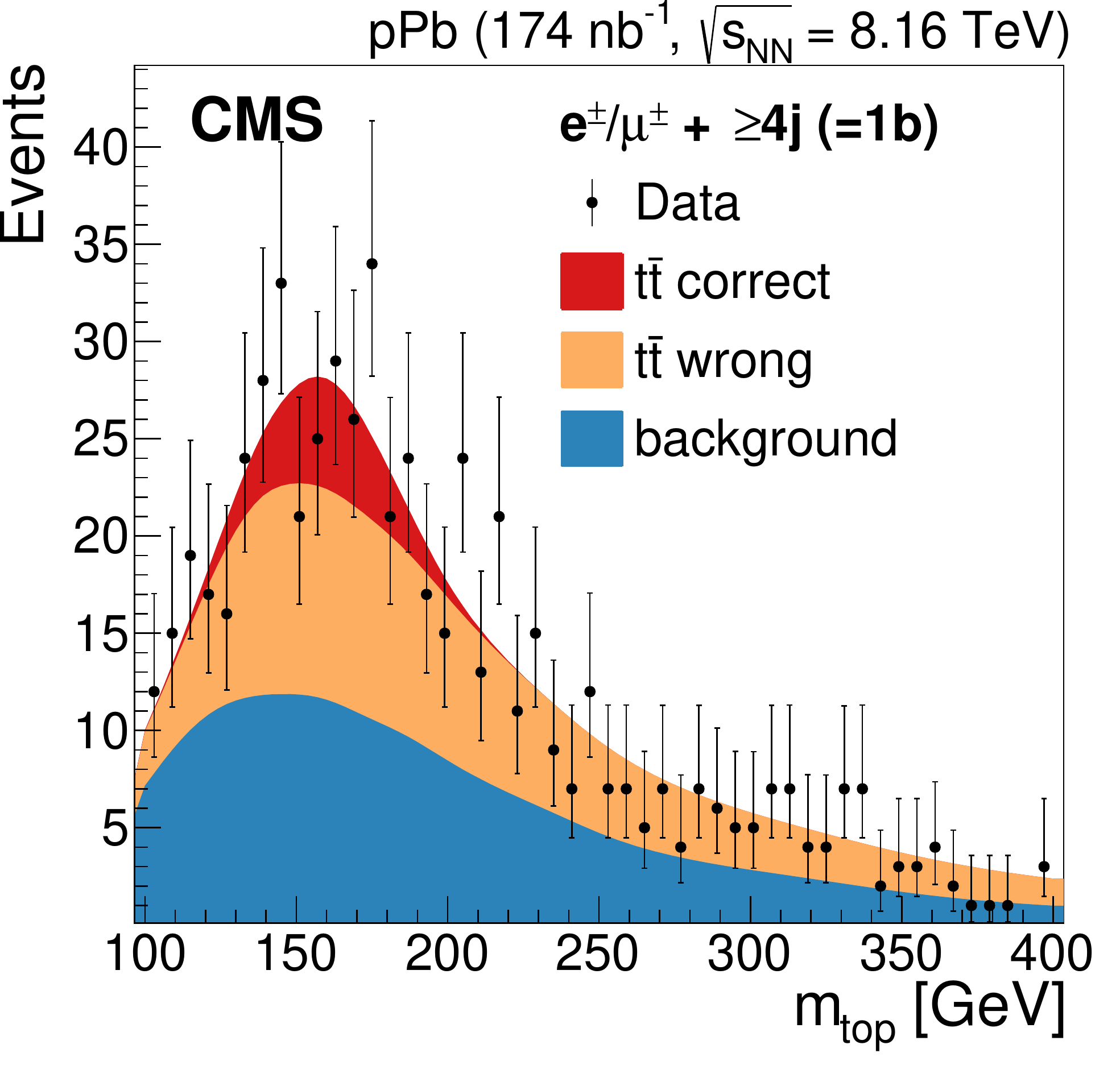}
\includegraphics[width=0.32\textwidth]{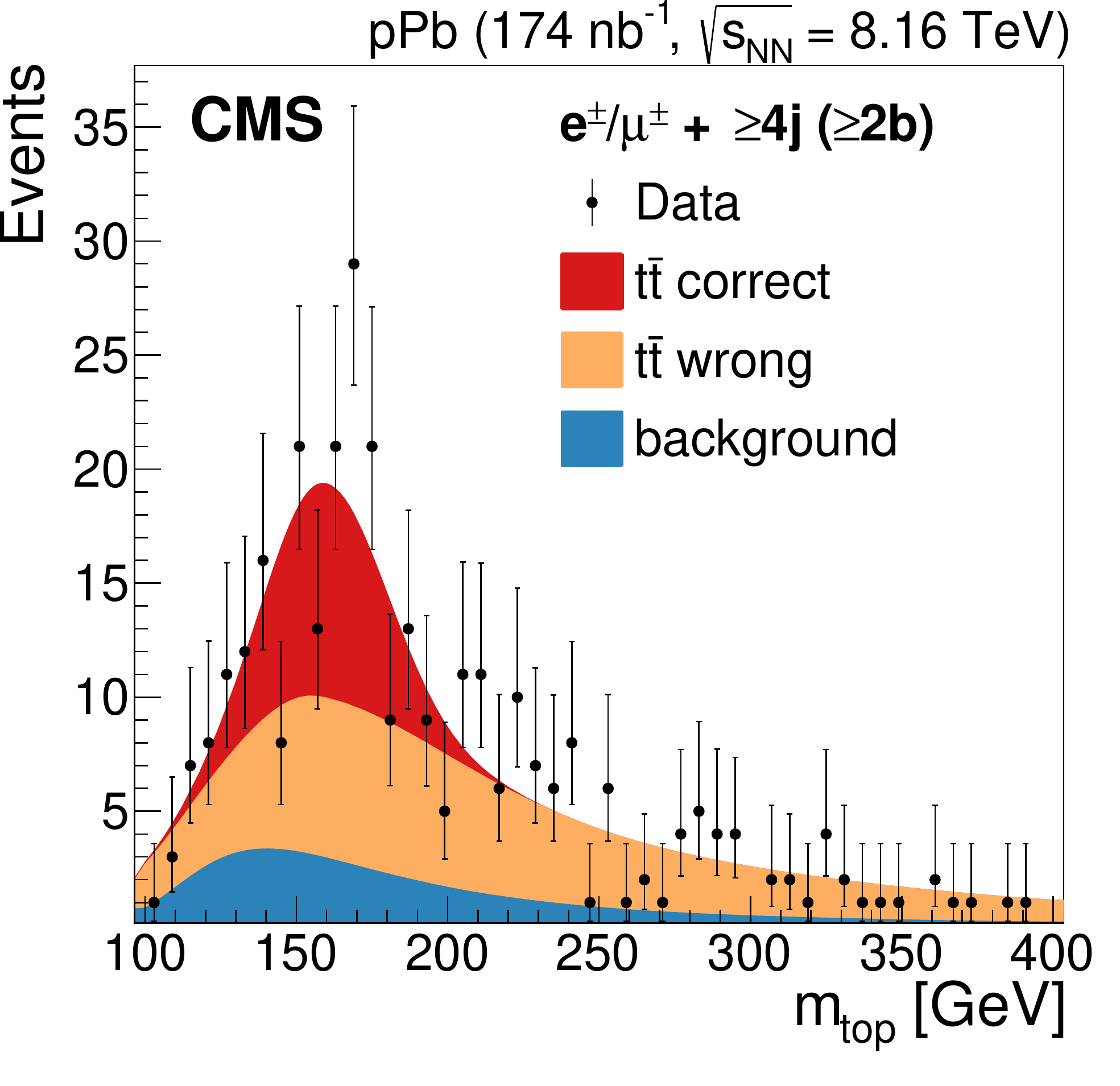}
\caption{Invariant mass distributions of the $\cPqt \to \jj\cPqb$ candidates, \mtop,
in the 0 (left), 1 (center), and 2 (right) \cPqb{}-tagged jet categories after all selections.
All signal and background parameters are kept fixed to the outcome of the \mjj fit.
Symbols and patterns are the same as in Fig.~\ref{fig:fitres}.}
\label{fig:valid-mthad}
\end{figure*}

To further examine the hypothesis
that the selected data are consistent with the production of top quarks, we define a proxy of the top quark mass, \mtop, as the invariant mass of a $\cPqt \to \jj \cPqb$ candidate formed by pairing the \PW\ candidate with a \cPqb-tagged jet. This pairing is chosen to minimize the absolute difference between the invariant masses of the $\cPqt \to \jj\cPqb$, and the $\cPqt \to \ell \nu \cPqb$ candidates.
In the 0b and 1b categories, the jet(s) with the highest value(s) of the \cPqb\ quark identification discriminator are considered for this purpose.
Figure~\ref{fig:valid-mthad} shows the distribution of \mtop reconstructed for events in the 0, 1, and 2 \cPqb-tagged jet categories,
with all signal and background parameters kept fixed to those from the outcome of the \mjj fit.

The total number of \ttbar\ signal events obtained through the fit of the $\mu$+jets and \Pe+jets channels combined, is 710.
Sources of experimental uncertainty in the measurement include the uncertainty in the \cPqb\ tagging efficiency, which is measured {\it in situ}
and bears the largest effect of $\pm$13\% on the \ttbar\ cross section; and the jet energy scale~\cite{Chatrchyan:2011ds}, which takes into account a
3\%-level difference between the reconstructed and generated jet energy in MC events and a 3\% residual calibration uncertainty from data, that together propagate as an
additional $\pm$4\% uncertainty in the final cross section.
Background shape and normalization uncertainties are also determined in the fit procedure
and have a $\pm$7\% effect on the extracted cross section. Uncertainties in the lepton trigger and reconstruction efficiencies, estimated with the tag-and-probe method,
result in a $\pm$4\% effect on the measured cross section.
The integrated luminosity calibration for \pPb data taking conditions results in a $\pm$5\% uncertainty.
The jet energy resolution~\cite{Chatrchyan:2011ds}, as estimated in proton-proton collision data, and the 0.1\% uncertainty of
 the LHC beam energy~\cite{Wenninger:2254678}, have a numerically insignificant effect on this measurement.

The compatibility of the data with a background-only hypothesis has been evaluated using
a profile-likelihood ratio as a test statistic~\cite{HiggsCombine} including all systematic uncertainties as nuisance parameters
with Gaussian priors.
Several tests have been performed, varying the estimation method and the background modeling assumptions.
Even with the most conservative assumptions, the background-only hypothesis is excluded with a significance above 5 standard deviations.
The \ttbar production cross section is then obtained via
\begin{linenomath}
\begin{equation}
\stt=\frac{S}{\mathcal{A}\,\varepsilon\,\mathcal{L}}\ ,
\label{eq:xsec}
\end{equation}
\end{linenomath}
where $S$ is the number of fitted signal events; $\mathcal{A} = 0.060 \pm 0.002$ and $0.056 \pm 0.002$ are the total acceptances in
the $\mu$+jets and \Pe+jets channels relative to all generated \ttbar events,
including the branching fraction to leptons, as determined from simulation;
$\varepsilon = 0.91 \pm 0.04$ and $0.63 \pm 0.03$ are the  $\mu$+jets and \Pe+jets event selection efficiencies as estimated from data;
and $\mathcal{L}$ is the total integrated luminosity. The 4\% uncertainty in the acceptance correction $\mathcal{A}$, including its
dependence on the proton and Pb PDFs, and on the values of theoretical scales and the QCD coupling
($\alpha_s = 0.118 \pm 0.001$ at the \PZ\ boson pole mass), has been determined
from a NLO $\pPb \to \ttbar+X$ sample generated with \POWHEG~(v.2)~\cite{powheg,powheg2,Frixione:2007nw}.
The total uncertainty on $S$ is obtained from the covariance matrix of the fit. It is further split into a statistical part,
by leaving \stt to float in the fit and fixing all other parameters to their post-fit values,
and a systematic part, by subtracting the square of the statistical uncertainty from the square of the total uncertainty.
From Eq.~(\ref{eq:xsec}), we measure
\begin{equation}\begin{aligned}
\stt^{\mu+\text{jets}}&=44\pm 3\stat\pm 8\syst\unit{nb},\\
\stt^{\Pe+\text{jets}}&=56\pm 4\stat\pm 13\syst\unit{nb},
\end{aligned}
\end{equation}
in the individual $\mu \text{+jets}$ ($S=420$) and $\Pe \text{+jets}$ ($S=348$) channels, with a relative total uncertainty of 18\% and 23\%, respectively.
The combined fit to both channels yields
\begin{linenomath}
\begin{equation}
\stt=45\pm 8\,\text{(total)}\unit{nb}.
\end{equation}
\end{linenomath}

\begin{figure}[htp]
\centering
\includegraphics[width=\cmsFigWidth]{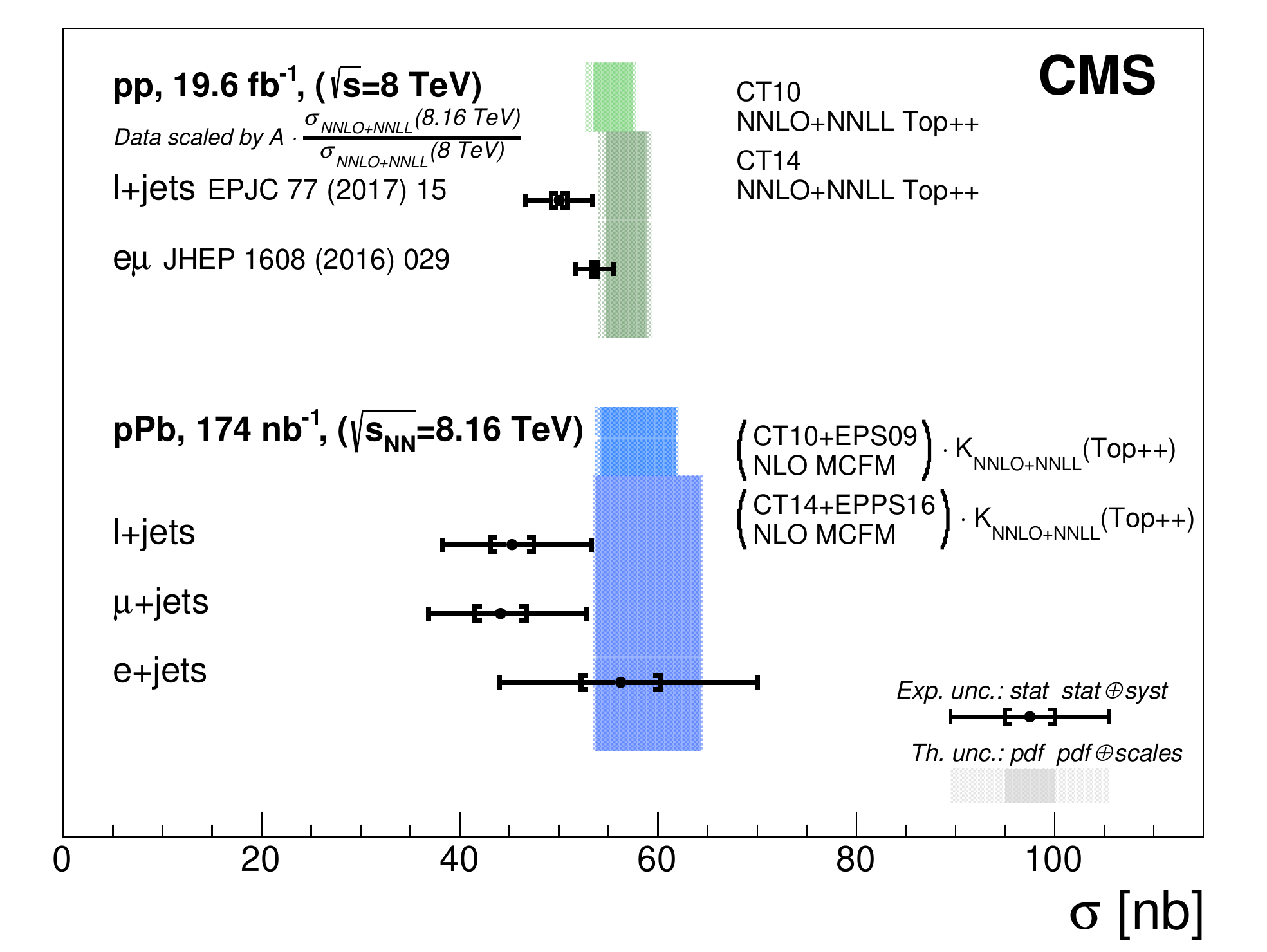}
\caption{Total \ttbar\ cross sections measured in the \Pe+jets, $\mu$+jets, and combined $\ell$+jets channels
in \pPb\ collisions at $\rootsNN=8.16$\TeV, compared to theoretical NNLO+NNLL predictions, and to scaled $\roots=8$\TeV pp results~\cite{Khachatryan:2016mqs,Khachatryan:2016yzq}.
The total experimental error bars (theoretical error bands) include statistical and systematic (PDF and scale) uncertainties added in quadrature.}
\label{fig:summary}
\end{figure}

The measured cross section is found to be consistent with the theoretical prediction~\cite{dEnterria:2017dum},
$\sigma(\pPb\to\ttbar + X) = 59.0 \pm 5.3\,(\mathrm{PDF})\,^{+1.6}_{-2.1}\,\text{(scale)}\unit{nb}$,
computed with \mcfm (v.8)~\cite{mcfm} using the CT14 proton PDF~\cite{ct14} and the EPPS16 nPDF for the lead
ions~\cite{EPPS16}, scaled to NNLO+NNLL accuracy with a $K$ factor computed with \textsc{Top++} (v.2.0)~\cite{top++},
and multiplied by $A = 208$.
The PDF uncertainties are obtained from the corresponding 56+40 eigenvalues of the CT14+EPPS16 sets
(corresponding to a 90\% confidence level) added in quadrature, while the theoretical scale uncertainty
is estimated by modifying the factorization and renormalization scales within
a factor of two with respect to their default value set at $\mu_\mathrm{F} = \mu_\mathrm{R} = m_{\cPqt}$.
The same calculation with the CT10 proton PDF~\cite{ct10} and EPS09~\cite{EPS09} nPDF yields
$\sigma(\pPb\to\ttbar+X) = 57.5 \pm \,^{+4.3}_{-3.3}(\mathrm{PDF})\,^{+1.5}_{-2.0}\,\text{(scale)}\unit{nb}$.
The difference in the theoretical \ttbar cross section computed with the PDF for free protons and for bound nucleons is small. A net overall
antishadowing effect increases the total top-quark pair cross section by only 4\% for both the EPPS16 and EPS09 sets in \pPb relative to \pp collisions~\cite{dEnterria:2017dum}.
Such a difference is too small to be observed in the data with the current experimental uncertainties.
Figure~\ref{fig:summary} shows the measured and
theoretical cross sections for \ttbar production in \pPb collisions at $\rootsNN=8.16$\TeV, compared with the results from \pp
collisions at $\roots=8$\TeV~\cite{Khachatryan:2016mqs,Khachatryan:2016yzq} scaled by $A$ and by the ratio of 8.16\TeV over 8\TeV NNLO+NNLL cross sections.

In summary,
the top pair production cross section has been measured for the first time in proton-nucleus collisions, using \pPb data at $\rootsNN = 8.16\TeV$ with a total integrated luminosity of 174\nbinv.
The measurement is performed by analyzing events with exactly one isolated electron or muon and at least four jets.
The significance of the \ttbar\ signal against the background-only hypothesis is above five standard deviations.
The measured cross section is $\stt=45\pm 8$\unit{nb}, consistent with the expectations from scaled \pp data as well as perturbative quantum chromodynamics
calculations.
This first measurement paves the way for further detailed investigations of top quark production in nuclear interactions,
providing in particular a new tool for studies of the strongly interacting matter created in nucleus-nucleus collisions.

\begin{acknowledgments}
We congratulate our colleagues in the CERN accelerator departments for the excellent performance of the LHC and thank the technical and administrative staffs at CERN and at other CMS institutes for their contributions to the success of the CMS effort. In addition, we gratefully acknowledge the computing centers and personnel of the Worldwide LHC Computing Grid for delivering so effectively the computing infrastructure essential to our analyses. Finally, we acknowledge the enduring support for the construction and operation of the LHC and the CMS detector provided by the following funding agencies: BMWFW and FWF (Austria); FNRS and FWO (Belgium); CNPq, CAPES, FAPERJ, and FAPESP (Brazil); MES (Bulgaria); CERN; CAS, MoST, and NSFC (China); COLCIENCIAS (Colombia); MSES and CSF (Croatia); RPF (Cyprus); SENESCYT (Ecuador); MoER, ERC IUT, and ERDF (Estonia); Academy of Finland, MEC, and HIP (Finland); CEA and CNRS/IN2P3 (France); BMBF, DFG, and HGF (Germany); GSRT (Greece); OTKA and NIH (Hungary); DAE and DST (India); IPM (Iran); SFI (Ireland); INFN (Italy); MSIP and NRF (Republic of Korea); LAS (Lithuania); MOE and UM (Malaysia); BUAP, CINVESTAV, CONACYT, LNS, SEP, and UASLP-FAI (Mexico); MBIE (New Zealand); PAEC (Pakistan); MSHE and NSC (Poland); FCT (Portugal); JINR (Dubna); MON, RosAtom, RAS, RFBR and RAEP (Russia); MESTD (Serbia); SEIDI, CPAN, PCTI and FEDER (Spain); Swiss Funding Agencies (Switzerland); MST (Taipei); ThEPCenter, IPST, STAR, and NSTDA (Thailand); TUBITAK and TAEK (Turkey); NASU and SFFR (Ukraine); STFC (United Kingdom); DOE and NSF (USA). \end{acknowledgments}
\bibliography{auto_generated}

\cleardoublepage \appendix\section{The CMS Collaboration \label{app:collab}}\begin{sloppypar}\hyphenpenalty=5000\widowpenalty=500\clubpenalty=5000\input{HIN-17-002-authorlist.tex}\end{sloppypar}
\end{document}

%% file: HIN-17-002-authorlist.tex
\textbf{Yerevan Physics Institute,  Yerevan,  Armenia}\\*[0pt]
A.M.~Sirunyan, A.~Tumasyan
\vskip\cmsinstskip
\textbf{Institut f\"{u}r Hochenergiephysik,  Wien,  Austria}\\*[0pt]
W.~Adam, F.~Ambrogi, E.~Asilar, T.~Bergauer, J.~Brandstetter, E.~Brondolin, M.~Dragicevic, J.~Er\"{o}, A.~Escalante Del Valle, M.~Flechl, M.~Friedl, R.~Fr\"{u}hwirth\cmsAuthorMark{1}, V.M.~Ghete, J.~Grossmann, J.~Hrubec, M.~Jeitler\cmsAuthorMark{1}, A.~K\"{o}nig, N.~Krammer, I.~Kr\"{a}tschmer, D.~Liko, T.~Madlener, I.~Mikulec, E.~Pree, N.~Rad, H.~Rohringer, J.~Schieck\cmsAuthorMark{1}, R.~Sch\"{o}fbeck, M.~Spanring, D.~Spitzbart, W.~Waltenberger, J.~Wittmann, C.-E.~Wulz\cmsAuthorMark{1}, M.~Zarucki
% ...
\vskip\cmsinstskip
% ...
1:~~Also at Vienna University of Technology, Vienna, Austria\\
% ...
